\documentclass[11pt,twoside]{article} 
\usepackage{acta-info}
\usepackage{euler}
\usepackage{multirow}
\usepackage{array}
\usepackage{booktabs}
\usepackage[section]{placeins}
\usepackage{float}
\usepackage{epstopdf}

\usepackage{graphicx}

\newcommand{\vol}{\textbf{x,} }   
\newcommand{\amss}{\amssubj{%
68U20
}}     
\newcommand{\CRs}{\CRsubj{%
F.1.3
}}   
\newcommand{\keyw}{\keywords{%
statistical complexity, Shannon entropy,   chaos, snapshot attractor, on-off intermittency 
}}

\begin{document}

\title{%
 Statistical complexity of the quasiperiodical  damped systems
}

\maketitle


\href{http://www.domain.edu}{\'Agnes F\"UL\"OP} \\  
\href{http://www.domain.edu}{E\"otv\"os University}\\ 
\href{mailto:fulop@caesar.elte.hu}{fulop@caesar.elte.hu} 


\begin{abstract}
We consider the concept of statistical complexity to write the quasiperiodical damped systems applying the snapshot attractors. This allows us to understand the behaviour of these dynamical systems by the pro\-bability distribution of the time series making a difference between the regular, random and structural complexity on finite measurements. We interpreted the statistical complexity on snapshot attractor and determined it on the quasiperiodical forced pendulum.   
\end{abstract}


\section{\bf Introduction}

 There is no universal definition of complexity in natural sciences. In the last two decades several complexity measures have been introduced, which contain various aspects of complex systems. We mention some ones as  algorithmic complexity (Kolmogorov) \cite{ank}, amount of information about the optimal predict the future corresponding to the expected past  (Crutchfield, Young) \cite{cy}, (Boffetta, Cencini, Falconi, Vulpiani)\cite{av}, complexity of finite sequence (Lempel, Ziv) \cite{lz}.
 
  An important contribution to this issue  due to P. Grassberger \cite{gpijt}, he stu\-died  the pattern-generation by the dynamics of a system and introduced the effective entropy considering the mixture of order and disorder, regularity and randomness, because the most complex situation is neither the one with highest Shannon information $\mathcal{S}$ (random structure) nor the one with lowest $\mathcal{S}$ (ordered structures). 
 
We negotiate the statistical complexity  in this article, which allows  a description of a finite measured series to consider  more complicated dynamical structures which was published  in the article \cite{lmc} (1995).  It  was widely used  in the chaotic regim  \cite{fpp}, biology \cite{saun}, symbolic sequences\cite{ac}, pseudorandom bit generator\cite{cghl}, earthquake \cite{lo}, the number system \cite{afs}. 

The entropy appoints the direction of flow. The complexity  determines the inner structure of the dynamical system. The statistical approaches are easier to implement than solving equations of motion and  they offer the only way of dealing with otherwise intractable problems.

We study the complexity of the quasiperiodic driven dynamical systems considering the snapshot attractor\cite{romeiras} on the set of points, which are determined by the Poencare section.  
This object corresponds to a given time moment, which  contains the points of the trajectories  ensemble. These orbits were initialized in the past and the time dependent behaviour was determined by the same equation of motion.

We determined the numerical approximation of the aperiodic driven pendulum, which reflects the behaviour of complexity on the snapshot attractor.  This system shows on-off intermittency\cite{lg} near to the axis (L=0), due to the fluctuation of the maximal Lyapunov exponent. The maximal Lyapunov exponent linearly intersects the axis, therefore it can be seen, that the system has scaling behaviour  \cite{scaling}. 

The structure of the article contains the next parts: 
In the section 1 we introduce the idea of complexity accordingly the measure of entropy and disequilibrium. We discuss the statistical complexity is extended to generalized complexity considering the Tsallis, Wooters, Renyi  entropy and the Kullbac-Shannon, Kullbac-Tsallis, Kulback-Renyi divergency. We explained the quasi periodical motion by snapshot attractor,  which disposes the on-off intermittency between chaotic and nonchaotic condition and this system provides the scaling behaviour in the section  2.  A numerical approximation of the aperiodic forced system  is illustrated by the quasiperiodic driven pendulum comparing the complexity and the dissipation rate, which due to the on-off intermittency.

\section{Complexity}
  
We investigate the statistical complexity, which is based on the  effective entropy by P. Grassberger\cite{gpijt} and the main idea is published by   R. L\'opez-Ruiz, H.L. Manchini, X. Calbet  (LMC) \cite{lmc,mpr,ap,cl}.

 The expanded definition of the statistical complexity so called generalized statistical complexity measures were introduced   by M.T. Martin, A. Plastino, O.A. Rosso (2006) \cite{mpo}, which apply various  kinds of entropy and disequilibrium measure  \cite{entr}. 

 We use  the notation of a  measured sequence by the article \cite{af2}. The time series of the ensemble is denoted by  $y_{1,j},\dots ,y_{n,j}$, where $y_{i,j}$ means the measurement of the quantity $y$ at time $t_i=t_0+i\Delta t$,  the time interval  $\Delta t >0 \in \mathbb{R}$ and $j$ $(1<j<m)$  assigned a number of the trajectory  in the ensemble. The unit of the time interval $\Delta t$ equals to a constant  in this description. The $\underline{x}^{(n)}$  indicates the trajectory of length $n$ in $\mathbb{R}^d$, which means a time series of the measurement. The $k$th point of the orbit of length $n$ in the manifold is  written by  ${{x}}^{(n)}_{k,j}$, where  $(k=1,\dots n)$, and $j$ means the number of trajectory $(j=1,\dots m)$ in the manifold.
 We will research the sequent of  $x_{1,j}^{(n)}, x_{2,j}^{(n)},\dots ,x_{n,j}^{(n)}$  as a time series over the $j$th trajectory.
   Let us choose this time development quantity of the ensemble  at a given  moment $i=t'$, then
 we determine the probability distribution of the ${{x}}^{(n)}_{t',j}  (j=1,\dots m)$  over the    points of the trajectories of the manifold.

We rephrase this  notation by the  symbolic dynamics, because the concept of  comp\-lexity is much more universal idea than this question. We distinguish $M$ different value of the measurements. The points of the trajectories of length $n$  ${{x}}^{(n)}_{i,j}$ $(1<i<n)$ $(1<j<m)$ are noted by the symbol $o_{i,j}$, which is chosen from the set $\{1,\dots M\}$. Then the $j$th path of length $n$ in the ensemble corresponds to $O^{(n)}_j=(o_{1,j},o_{2,j},\dots o_{n,j})$. A series $O^{(n)}_j$ $(1<j<m)$  occurs with probability $P(O^{(n)}_j)$ along  a sequent of length $N$ ($n< N$).

\subsection{Statistical complexity}

The statistical complexity is well applicable concept characterizing finite measurement sequences with its probability distribution. This allows a statistical approximation of the measured quantities. We apply the basic article \cite{lmc} to introduce this idea.

We suppose that there are $N$ various symbol series of length $n$ $(o_{1,j}, \dots o_{N,j})$ $(j=1,\dots m)$ in the ensemble, then these series dispose the set of discrete probability distribution $P\equiv \{p_{1,j},\dots p_{N,j}\}$, where $p_{i,j}:=P(o_{i,j})$ $(\sum_{i=1}^Np_{i,j}=1)$ $(1\le j \le m)$ and $p_{i,j}>0$ for all $i$.

The statistical complexity measures contains two compositions: (i) entropy $\mathcal{H}$  and (ii) disequilibrium $\mathcal{D}$ i.e. distances in probability-space. It is introduced by the information theory, where the Shannon entropy assigns the gain of the information storage and the disequilibrium features the distance from uniform distribution. So the LMC measure gives a simultaneous quantification of randomness and correlation structures in the systems.

\subsubsection{Measure of Entropy and Disequilibrium}

\paragraph{Information measure}

Information measure $\mathcal{I}[P]$ i.e.  the uncertainty  can be  described by the probability distribution $P=\{p_j, j=1,\dots ,N\}$, with $N$  the number of possible states of the system ($\sum_{j=1}^N p_j=1$). In the information theory we define the quantity of disorder $\mathcal{H}$ for a given probability distribution $P$ corresponding to the information measure $\mathcal{I}[P]$ in the next term:
\begin{eqnarray}
\mathcal{S}[P]=\mathcal{I}[P]/\mathcal{I}_{max}[P_e],
\end{eqnarray}
where $0\le \mathcal{H} \le 1$, $\mathcal{I}_{max}$ means the maximal value of $\mathcal{I}$, and $P_e$ is the uniform probability distribution. Let us consider the Shannon-Kinchin paradigm  then  $\mathcal{I}$ can be defined as a term of entropies. The statistical complexity was introduced by the Shannon entropy \cite{cs}, therefore we will investigate this form in a finite system:
\begin{eqnarray}
\mathcal{S}=-\sum_{i=1}^N p_i \log p_i.
\end{eqnarray}
This quantity approximately equals to zero $\mathcal{S}\sim 0$, if the symbol sequence $O^{(n)}_{j_c}$ has a high probability ($p_c \sim 1$) and other $O^{(n)}_j$  has very small probability ($p_c \sim 0$). In the range of entropy the maximal values  $\mathcal{S}_{max}$ corresponds to the uniform probability dist\-ribution $p_e=\{ 1/N, 1/N, \dots 1/N\}$, which means the the equal probability symbol sequence $O_{j_e}^{(n)}$, which leads to the  maximum value of information.  The normalized quantity ${\mathcal{H}}$ derives from ${\mathcal{H}}=\mathcal{S}/\mathcal{S}_{max}$, than $0\le {\mathcal{H}}\le 1$, where $\mathcal{S}_{max}=\log N$.

\paragraph{Disequilibrium measure}

We introduce the function of disequilibrium $\mathcal{D}$ on the probability distribution $\{p_j: j=1\dots N\}$. The LMC uses some distance $\mathcal{D}$ of a given $P$ compared to the uniform distribution $P_e$ in the states of the system \cite{lmc}.
Therefore we investigate the disequilibrium by a distance-form:
\begin{eqnarray}
\mathcal{Q}[P]=\mathcal{Q}_0\cdot \mathcal{D}[P,P_e],
\end{eqnarray}
where $\mathcal{Q}_0$ is a normalization constant $(0\le \mathcal{Q}_0 \le 1)$, which equals to the inverse of the maximum possible value of the distance $\mathcal{D}[P,P_e]$. The maximum distance is obtained, when one of the components of $P$ i.e. $p_l$  equals to one and the remaining elements equal to zero. The disequilibrium $\mathcal{Q}$ differs from zero, if there exist preferred states among the accessible ones, i.e. this quantity features  the systems' architecture.

 In the original definition of the statistical complexity   the Euclidean norm  ($\mathbb{R}^N$)  have been used  i.e. the quadratic distances from the probability distribution of each symbol sequences $P(O^{(n)}_j)$ $(1\le j\le m)$ to the equal probability $P(O^{(n)}_{j_e})$. This choice for the distance $\mathcal{D}$ is written:
\begin{eqnarray}
 \mathcal{D}[P,P_e]=\sum_{i=1}^N \left(p_i-p_e\right)^2,\;\; \mbox{where} \;\; p_e=\frac{1}{N}.
 \end{eqnarray}
In the range of this quantity becomes maximum, when the disequilibrium achieves  prevalent symbol sequences $O^{(n)}_{j_c}$ with $p_c\sim 1$ and $\mathcal{D}_c \to 1$ for $N$ is growing. Otherwise the disequilibrium equals to zero approximately $\mathcal{D}\sim 0$ for $p_i \sim 1/N$. The value of the $\mathcal{D}$ changes between these extrema corresponding to advanced  probability distribution.
The normalized factor  equals to $\mathcal{Q}_0=\frac{N}{N-1}$.

\subsubsection{Measure of statistical complexity}\label{msc}

The whole complexity concept includes the  functional product of disorder ${\mathcal{H}}$ and disequilibrium $\mathcal{D}$, which based on the various probability distribution corresponding to sequent of the advanced quantity. This shows the transition between the information stored in the system and its disequilibrium. The statistical complexity $\mathcal{C}$ is introduced by the published article of LMC  \cite{lmc}: 
\begin{eqnarray}\label{lmc-m}
\mathcal{C}={\mathcal{H}}\cdot \mathcal{D}=-\left(\sum_{i=1}^N p_i \log p_i\right)\left(\sum_{i=1}^N\left(p_i-\frac{1}{N}\right)^2\right).
\end{eqnarray}
The value $\mathcal{C} \in \mathbb{R}^+$. The normalized  $\mathcal{C}$  can be described as follows $\overline{\mathcal{C}}=\overline{\mathcal{H}}\cdot \overline{\mathcal{D}}=(\mathcal{H}/ \log N)(\mathcal{D}\cdot(N/(N-1)))$. The range of complexity measure is finite and limiting between $\mathcal{C}_{min}$ and $\mathcal{C}_{max}$, but $\mathcal{H}$ is not necessarily a unique function.

We applied finite system, therefore the statistical complexity disposes the scaling behaviour. 
A new set of symbol sequences $O^{(n)}_j$ turns out at each scale of measurement, which provides an advanced probability distribution   $P(O^{(n)}_j)$ so the value of complexity becomes different.

Three basic cases are distinguished of the statistical complexity: (a) this is monotonous increasing in the function of entropy  (b) it corresponds to a convex function, which contains a maximal $\mathcal{C}_{max}$ at the probability distribution $p_e$ and the minimum $\mathcal{C}_{min}$ appears, where the   ${\mathcal{H}}=0$ i.e.  total order and ${\mathcal{H}}=1$ and
third kind is (c) the   monotonous decreasing with increasing  entropy \cite{mpo}. 

There are two extremist situations of  complexity $\mathcal{C}$ depending on entropy $\mathcal{H}$. On the one hand each set of series assigned to each set of symbol sequence $O^{(n)}_j$, which has the same probability distribution. All of them contribute to the information stored in equal measure as the ideal gas \cite{pl1}. On the other hand, If we study an object, which features  some symmetries properties and distance, then this system can be written by minimal information as mineral or symmetry in quantum mechanics.

\subsubsection{Generalized statistical complexity}

\paragraph{Generalized entropy}
We expend the concept of classical statistical complexity to different measures of the entropy and disequilibrium. Tsallis introduced a generalisation of the Shannon-Boltzmann-Gibbs entropic measure \cite{ct}:
\begin{eqnarray}
\mathcal{S}_q^{(T)}[P]=\frac{1}{(q-1)}\sum_{j=1}^N\left[p_j-(p_j)^q\right],
\end{eqnarray} 
where $q$ real number. Renyi suggested a definition of entropy  for discrete probability distribution in 1950s years \cite{ar}:
\begin{eqnarray}
\mathcal{S}_q^{(R)}[P]=\frac{1}{(1-q)}\ln \left\{\sum_{j=1}^N(p_j)^q\right\}. 
\end{eqnarray}
Then the generalized entropy  $\mathcal{S}_q^{( \kappa)}$ denotes  $\kappa=S,T,R$ Shannon,Tsallis and Renyi entropy. 
\begin{eqnarray}
\mathcal{H}_q^{(\kappa)}[P]=\mathcal{S}_q^{( \kappa)}[P]/\mathcal{S}_{max}^{(\kappa)},
\end{eqnarray}
where $\mathcal{S}_{max}^{(\kappa)}$ means the maximum value of the information measure, which corresponds the uniform probability distribution.  The maximal value of Shannon and Renyi entropy correspond to $\mathcal{S}_{max}^{(S)}=\mathcal{S}_{max}^{(R)}=\ln N$ and  Tsallis entropy involves  $\mathcal{S}_{max}^{(T)}=\frac{1-N^{1-q}}{q-1}$ for $q\in (0,1)\cup (1,\infty)$.   

\paragraph{Generalized disequilibrium}  
The Euclidean distances was investigated in the LMC. We introduce the generalized disequilibrium $\mathcal{D}$. The  Wootters distance was applied for two probability distributions   \cite{wkw}, which can be used in the quantum mechanic:
\begin{eqnarray}
\mathcal{D}_W[P_1,p_2]=\cos^{-1}\left\{\sum_{j=1}^N (p_j^{(1)})^{1/2} (p_j^{(2)})^{1/2} \right\}.
\end{eqnarray}
Consider two divergence-classes, which were published by Basseville \cite{bas}. The first class contains the divergence which is defined by the relative entropies. The second class consists of the divergence, which was introduced as the difference of the entropies. The Kullbac-Shannon expression  following
\begin{eqnarray}
\mathcal{D}_{\mathcal{K}^S}[P,P_e]=\mathcal{K}^{(\mathcal{S})}[P|P_e]=\mathcal{S}_1^{(\mathcal{S})}[P_e]-\mathcal{S}_1^{(\mathcal{S})}[P].
\end{eqnarray}
The Kullback-Tsallis entropy is introduced:
\begin{eqnarray}
\mathcal{D}_{\mathcal{K}^T_q}[P,P_e]=\mathcal{K}_q^{(T)}[P|P_e]=N^{q-1}(\mathcal{S}_q^{(T)}[P_e]-\mathcal{S}_q^{(T)}[P]).
\end{eqnarray}
The Kulback-Renyi etropy following
\begin{eqnarray}
\mathcal{D}_{\mathcal{K}^R_q}[P,P_e]=\mathcal{K}_q^{(R)}[P|P_e]=(\mathcal{S}_q^{(R)}[P_e]-\mathcal{S}_q^{(R)}[P]).
\end{eqnarray}
Then the generalized disequilibrium denoted by this form:
\begin{eqnarray}
\mathcal{Q}_q^{(\nu)}[P]=\mathcal{Q}_0^{(\nu)}\mathcal{D}_{\nu}[P,P_e],
\end{eqnarray}
where $\nu=E,W,K,K_q$ and $\mathcal{Q}_0^{(\nu)}$ normalization constant ($0\le \mathcal{Q}_q^{(\nu)}\le 1$), and these are: 
\[
\begin{array}{ll}
\mathcal{Q}_0^{(E)}=\frac{N}{N-1}, & \mathcal{Q}_0^{(W)}=1/\cos^{-1}\left\{\left(\frac{1}{N}\right)^{1/2}\right\},\\
\mathcal{Q}_0^{\mathcal{K}_q^R}=\frac{1}{\ln N}  & \mathcal{Q}_0^{\mathcal{K}_q^T}=\frac{q-1}{N^{(q-1)}-1}.
\end{array}
\]

\paragraph{Generalized statistical complexity}
This quantity is defined following 
\begin{eqnarray}\label{gsc}
\mathcal{C}_{\nu,q}^{(K)}[P]=\mathcal{Q}_q^{(\nu)}[P]\cdot \mathcal{H}_q^{(K)}[P],
\end{eqnarray}
where $K=S,R,T$ for fixed $q$ and the index $\nu=E,W,K_q$ means the disequilibrium with appropriated distance measures. This term (\ref{gsc}) is a family  of the statistical complexity corresponding to functional product form $\mathcal{C}=\mathcal{H}\cdot \mathcal{Q}$. We notice that the entropic difference $\mathcal{S}[P_1]-\mathcal{S}[P_2]$ does not  specify the information gain or divergence, because this quantity is not necessary positive definit. This lead to the Jensen divergence. 

Shinner, Davidson and Landsberg (SDL) published a term for the statistical complexity \cite{plx},
   this term is expressed for $\nu=K,K_q$:
\begin{eqnarray}
\mathcal{C}_{K_q}^{(\kappa)}[P]=(1-\mathcal{H}_q^{(\kappa)}[P])\cdot \mathcal{H}_q^{(\kappa)}[P].
\end{eqnarray}
We get the LMC statistical complexity at $\kappa=S$, $q=1$ so  $\mathcal{C}_{LMC}=\mathcal{C}_{K,1}^{(S)}$.

\subsection{\bf Time evolution}

In statistical physics we study the isolated systems, which are  characterized by  initial, arbitrary and discrete probability distribution. The uniform distribution $P_e$ develops during the evolution towards equilibrium. Then we can research the time evolution of the LMC i.e. $\mathcal{C}$ versus time $t$ graph. Thanks to the second law of thermodynamics the entropy grows monotonically with time ($d\mathcal{H}/dt\ge 0$) in isolated system. Therefore $\mathcal{H}$ behaves as an arrow of time, so we can study the figure of $\mathcal{C}$ versus $\mathcal{H}$ as the time evolution of the LMC, thus the normalised entropy-axis can be substituted by the time-axis. This picture $\mathcal{H}\times \mathcal{C}$ can be utilized to research the  changes in the dynamics of a system, which derives from the modulated parameters.

\section{\bf Driven system}\label{ds}

\vspace{0.3cm}

In both experimental and theoretical physics, periodically excited nonlinear systems play  important role. A typical case of these systems is described by this equation:
\begin{eqnarray}
\frac{d^2 \Theta}{dt^2}+ \nu \frac{d\Theta}{dt}+\sin \Theta = f(t),
\end{eqnarray}
where the damped forcing f(t) is periodic in time, for example:
\begin{eqnarray}
f(t)=K+V\cos(\omega t).
\end{eqnarray}
Such equations are used in many cases of physical research, as forced damped pendulum, the Stewart-McCumber model of the current-driven Josephson junction \cite{aa} and simple phenomenological model of sliding charge-density waves \cite{tb}. The nonlinear dynamical model can be characterized by strange attractor, period doubling cascades, crises, intermittency, fractal basin boundaries etc. These equations are intensively used in low dimensional chaotic dynamics research.

Periodic excitation is extended to examine aperiodic cases, when $f(t)$ is quasiperiodic for example:
\begin{eqnarray}
f(t)=K+V[\cos (\omega_1 t)+\cos(\omega_2 t)],
\end{eqnarray}
where $\omega_1$ and $\omega_2$ are the incommensurate frequencies. It appears in the transition from quasiperiodic to chaos inside an electronic  Josphson-junction simulator driven by two independent sources \cite{tb}. It is applied on the experiments inside an electron-hole plasma in germanium excited by two frequencies quasiperiodic external perturbations they observed stable three frequencies mode locking, and chaos \cite{3,4}.

These are presented with  quasiperiodic systems by two incommensurate frequencies. We consider the following quasiperiodically forced damped pendulum\cite{7}: 
\begin{eqnarray}\label{pend}
\frac{d^2 \Theta}{dt^2}+ \nu \frac{d\Theta}{dt}+\sin \Theta = K+V[\cos (\omega_1 t)+\cos(\omega_2 t)],
\end{eqnarray}
where $\Theta$ is an angle of pendulum with the vertical axis, $\nu$ is the dissipation rate, $K$ is a constant, $V$ is the forcing amplitude and $\omega_1$ and $\omega_2$ are the incommensurate frequencies. Let us investigate new variables, $t\to\nu t$ and $\phi=\Theta+\frac{\pi}{2}$. Then the equation (\ref{pend}) becomes:
\begin{eqnarray}
\frac{1}{p}\frac{d^2 \phi}{dt^2}+\frac{d\phi}{dt}-\cos \phi =K+V[\cos(\omega_1 t)+\cos(\omega_2 t)],
\end{eqnarray}
where $p=\nu^2$ is a new parameter, and $\omega_1$ and $\omega_2$ are rescaled as follows: $\omega_1 \to \omega_1\nu$ and $\omega_2 \to \omega_2 \nu$. In the expressions of the dynamical variables $\phi$,$v\equiv\frac{d\phi}{dt}$ and $z\equiv \omega_2 t$, then we have
\begin{eqnarray}\label{egyr}
\left.
\begin{array}{rcl}
\frac{d\phi}{dt}& =& v, \\
\frac{dv}{dt} & = & p\left\{ K+V\left[\cos\left(\frac{\omega_1}{\omega_2}z\right)+\cos z\right]+\cos \phi -v \right\},\\
\frac{dz}{dt} & = & \omega_2.\\
\hline
\end{array}
\right\}
\end{eqnarray}
The equation (\ref{egyr}) contains rich dynamical behaviour \cite{dp}. This system shows a special behaviour in some range of parameter space. Therefore we apply the snapshot attractor, because this structure reflects the properties of a dynamical systems at a given moment.

The dynamic of the quasiperiodic damped pendulum is characterised by the sign of maximal Lyapunov exponent which changes near to the axis (L=0), because this system has finite fluctuation.  
The Lyapunov exponent becomes to negative, then the system  contracts on the nonchaotic side($L\le 0$). Otherside  the Lypunov exponent turns into positive, then the expansion characterizes the dynamical behaviour on the chaotic side ($L \ge 0$). Therefore  the  collective properties of the orbits can be studied near to the transition. This is a special behaviour of this model which is called on-off intermittency of the snapshot attractor. The trajectories spend stretches of time expanding (leading to nonzero-size snapshot attractor), yet there are also long time during  the trajectories experience contraction, resulting extremely small-size of snapshot attractor. Then the time dependent size of snapshot attractor can be written by the dispersion rate \cite{iak}:

\begin{eqnarray}\label{st}
S(t)=\left(\frac{1}{N}\sum_{i=1}^N \left\{[\phi_i(t)-<\phi(t)>]^2+[v_i(t)-<v(t)>]^2\right\}\right)^{1/2},
\end{eqnarray}
where $N$ is the number of points on the snapshot  attractor, $[\phi(t),v(t)]$ defines the geometric center of these points at a given time: $<\phi(t)>=\frac{1}{N}\sum_{i=1}^N\phi_i(t)$ and $<v(t)>=\frac{1}{N}\sum_{i=1}^Nv_i(t)$.

The time averaged size of the snapshot  attractor on the chaotic side near to the transition is defined as $<S(t)>=\lim_{T\to \infty}\int_0^T S(t)dt$ which obeys the following scaling relation:
\begin{eqnarray}\label{scal}
<S(t)>\sim L \sim |p-p_c|,
\end{eqnarray}
where $p_c$ means the $p\ge p_c$ ($L \le 0$) and $p\le p_c$ ($L \ge 0$). 
 This scaling behaviour of the transition to chaos was published in quasiperiodically driven dynamical systems\cite{scaling} i.e. a route of chaos was investigated, where the largest Lyapunov exponent passes through zero linearly near the transition to chaos. Because the orbits  burst out to separate from each others during the expansion time intervals \cite{ps}, and the trajectories merge all together  during the contraction time interval therefore the size of the chaotic snapshot attractor changes widely in time near to the transition in an intermittent fashion. The average size of the snapshot   attractor scales linearly with a parameter similarly as the average interval between the bursts also scales linearly with parameter during transition (\ref{scal}).

\section{\bf Numerical approximation}

In this section we consider the statistical complexity of the quasiperiodic driven systems, which is represented by 
the aperiodic forced pendulum (\ref{pend}). We calculated the Poencare section ($z=0$) of the time dependent model (\ref{egyr}), which contains a snapshot attractor  at the parameter values $V=0.55,p=0.6, K=0.8$ $\omega_1=(\sqrt{5}-1)/2$, $\omega_2=1.0$. The initial values of the orbits are chosen by uniform distribution in a small volume $<< 10^{-6}$ in the phase space.  
We show the predictability of the intermittency between chaotic and nonchaotic regions. 

\paragraph{Statistical complexity}

We determined the statistical complexity of this system which  was introduced by the quantity of information theory, where the entropy and the disequilibrium depend on the  probability distribution (Section\ref{msc}).
The snapshot attractor is written by the ensemble of trajectories  instead of the long orbit $N'$. Therefore we  redefine the probability distribution of the manifold at a time instant $t'$. 

The ensemble of the snapshot attractor  contains the $x_{1,j}^{(n)}, x_{2,j}^{(n)},\dots ,x_{n,j}^{(n)}$  points of the trajectories of length $n$ $(1\le j \le m)$.
Therefore the $x_{t',1}^{(n)}, x_{t',2}^{(n)},\dots ,x_{t',m}^{(n)}$ series corresponds to $\{p_1,p_2\dots p_m\}$ probability distribution, where $p_j:=P(x_{t',j}^{(n)})$ ($j=1,\dots m$). 
   
\begin{figure}
\begin{center}
\includegraphics[width=8.0cm]{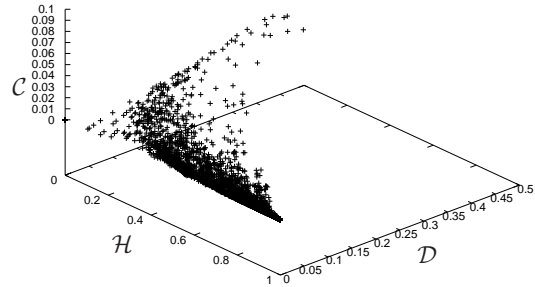}
\scalebox{0.85}{\put(-80,30){\makebox(0,0)[t]{ $\mathcal{D}$}}}
\scalebox{0.85}{\put(-220,35){\makebox(0,0)[t]{ $\mathcal{H}$}}}
\scalebox{0.85}{\put(-270,105){\makebox(0,0)[t]{ $\mathcal{C}$}}}
\caption{The complexity $\mathcal{C}$ depends on the entropy $\mathcal{H}$ and disequilibrium $\mathcal{D}$ for different parameter $p$ ($0<p<1$).}\label{hdc}
\end{center}
\end{figure}

The entropy $\mathcal{H}$, disequilibrium  $\mathcal{D}$ and statistical complexity $\mathcal{C}$ can be  determined by appropriate measures using the term of LMC complexity (\ref{lmc-m}).   
The structure of the complexity is plotted in the $\mathcal{C}\times \mathcal{H}\times \mathcal{D}$ space (Figure \ref{hdc}), where parameter $p$ changes between 0 and 1. The behaviour of complexity $\mathcal{C}$ monotone decreasing, so this corresponds to class (c) over the parameter range [0,1]. In a small interval at $p\simeq 0.8$ the shape of complexity formed convex curve (class (b)).

\paragraph{On-off intermittency}
    
\begin{figure}
\begin{center}
\includegraphics[width=6.0cm]{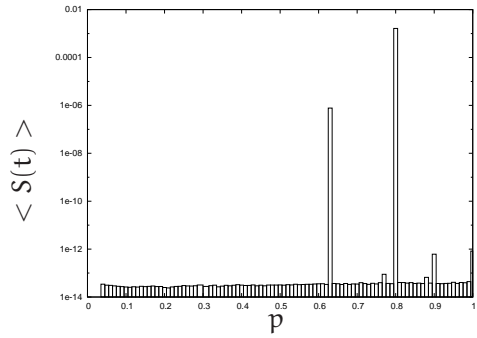}
\scalebox{0.95}{\put(-90,0){\makebox(0,0)[t]{ ${p}$}}}
\scalebox{0.95}{\rotatebox{90}{\put(70,205){\makebox(0,0)[t]{ ${<S(t)>}$}}}}
\caption{The  $<S(t)>$ depends on parameter $p$ in logarithm scale.}\label{2}
\end{center}
\end{figure}

\begin{figure}
\begin{center}
\includegraphics[width=10.0cm]{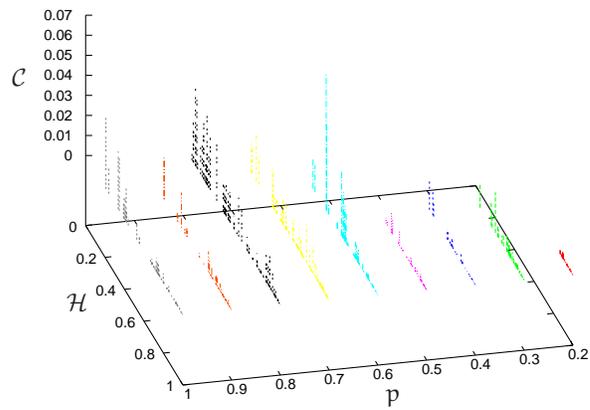}
\scalebox{0.85}{\put(-145,27){\makebox(0,0)[t]{ $p$}}}
\scalebox{0.85}{\put(-290,70){\makebox(0,0)[t]{ $\mathcal{H}$}}}
\scalebox{0.85}{\put(-320,170){\makebox(0,0)[t]{ $\mathcal{C}$}}}
\caption{The complexity $\mathcal{C}$ depends on the entropy $\mathcal{S}$ and parameter $p$.}\label{3}
\end{center}
\end{figure}

The quasiperiodic damped pendulum provides on-off intermittency in a special values of parameter, where  Lyapunov exponent has a finite fluctuation around axis of $L=0$ as it was detailed in the section  (\ref{ds}).
    The volume of the snapshot attractor is widely changing near to axis ($L=0$), therefore we applied    
   the  dispersion rate $S(t)$ (\ref{st}), which scales by the maximal Lyapunov exponent (\ref{scal}).
The structure of the complexity $\mathcal{C}(p)$  shows local maximums (Figure \ref{3}) similarly as the average of the dispersion $<S(t)>$  over the parameter space $p$ (Figure \ref{2}). 
The complexity reflects the behaviour of the pendulum i.e. the on-off intermittency. 
 Then the probability distribution of the snapshot attractor i.e. dispersion of the trajectories' points in the manifold at a given time instant $t'$  shows similar behaviour as the affect of the on-off intermittency.

\section{\bf Conclusion}

The inner structure of statistical complexity is determined in a quasiperiodical driven system at a given scale. The effect of the  on-off intermittency appears in complexity of the aperiodic damped system, which allows the predictability by the by the distribution probabililty of the snapshot attractor.

\end{document}